# EFFECTIVE PERFORMANCE OF INFORMATION RETRIEVAL ON WEB BY USING WEB CRAWLING

Sk.AbdulNabi[1] and Dr. P. Premchand[2]

[1]Department of CSE, AVN Inst. of Engg. & Tech, Hyderabad, A.P., India
nabi.cse@gmail.com
[2]Department of CSE, Osmania University, Hyderabad, A.P., India
drpremchand_p@yahoo.com

## ABSTRACT

*World Wide Web consists of more than 50 billion pages online. It is highly dynamic [6] i.e. the web continuously introduces new capabilities and attracts many people. Due to this explosion in size, the effective information retrieval system or search engine can be used to access the information. In this paper we have proposed the EPOW (Effective Performance of WebCrawler) architecture. It is a software agent whose main objective is to minimize the overload of a user locating needed information. We have designed the web crawler by considering the parallelization policy. Since our EPOW crawler has a highly optimized system it can download a large number of pages per second while being robust against crashes. We have also proposed to use the data structure concepts for implementation of scheduler & circular Queue to improve the performance of our web crawler. (Abstract)*

## 1. INTRODUCTION

In world wide web, information retrieval system [1] acts a vital role to extract the useful information with minimum amount of time. It has the ability to store the information and retrieved the information from the repository and also manages the information. The general objective of any Information Retrieval Method is to minimize the overhead of a user locating needed information.

The success rate information retrieval is depends on which information is required and what percentage of overhead is accepted by the user. In this context needed (required) information may be defined as adequate information in the system to complete a job. For example an Employee is willing to purchase a house from the realtor and then he needs all relevant information about the site and house documents and also previous history of site / house, where as student requires only limited information to success their examinations.

Some times a system needs smaller amount features instead of complete retrieval. In some cases complete retrieval is a negative feature because it overloads the user then it takes much time to get relevant information. This makes it more difficult for the users to filter the relevant but non-useful information from the critical items. Defining the "Relevant " term is faintly changed in system point of view than the user's point of view. Relevant is represents the needed information. From the user's viewpoint "relevant" and "needed" are identical, but from the system point of view information may be relevant to a search statement even though it is not needed / relevant to user.

Precision and recall [2] measures are plays very important role in the information system. When the user starts the searching information, then the database will be logically separated in to 4 parts. There are Relevant Retrieved, Non Relevant Retrieved, Relevant Not Retrieved and Non Relevant Not Retrieved shown in figure 1. Relevant items are those documents that contain information that helps the searcher in answering his question. Precision and recall are defined as:

Precision = Number_Retrieved_Relevant /
                    Number_Total_Retieved.
Recall     = Number_Retrieved_Relevant /
                    Number_Possible_Relevant

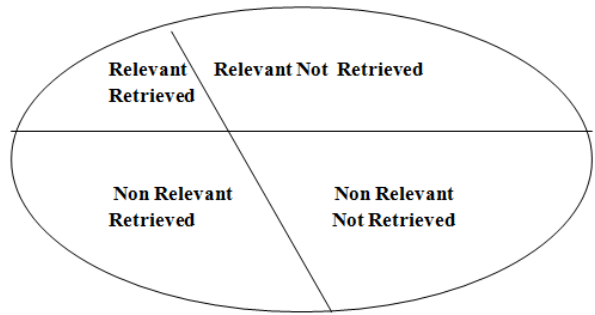

Figure 1.    Search on Segments of  Total Document Space

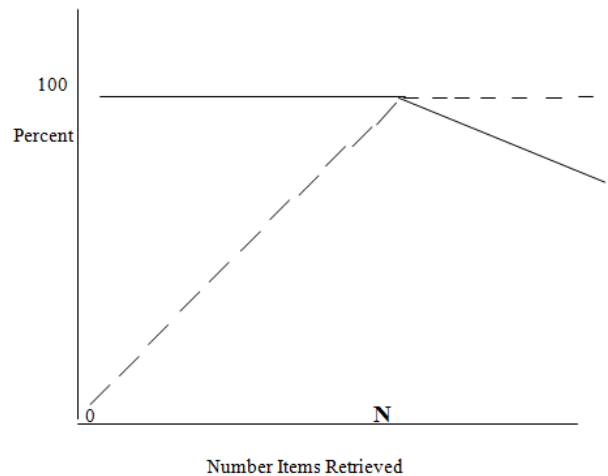

Figure 2. Ideal Precision and Recall

Where Number_Possible_Relevant are the number relevant items in the database. Number _ Total _ Retrieved is the total number of items retrieved from the query. Number _ Retrieved _ Relevant is the number of items retrieved that are relevant to the users search need.  If a search has a 85% precision, then 15% of the user effort is overhead reviewing non-relevant items. Recall represents the total number of relevant information retrieved from total number of possible relevant information. The relation ship between precision and recall has been shown in figure 2.

Figure.3 and 4 shows the optimal relationship between Precision and Recall. Precision starts of at 100% and maintains that values as long as relevant items are retrieved. Once all "N" relevant items have been retrieved, the remaining items been retrieved are Non-relevant. As with figure.2, in the ideal case every item retrieved is relevant. Thus precision stays at 100% (1.0).

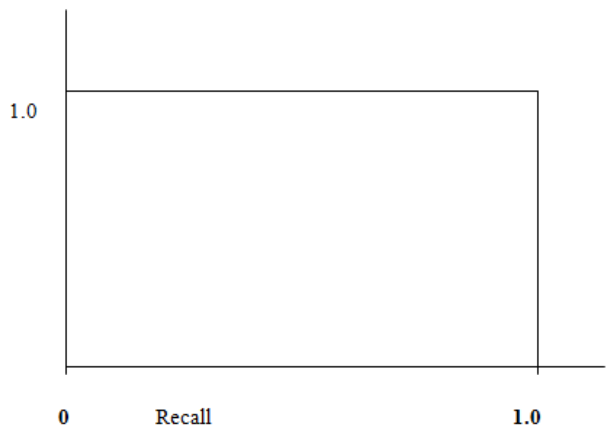

Figure 3. Ideal Precision / Recall graph

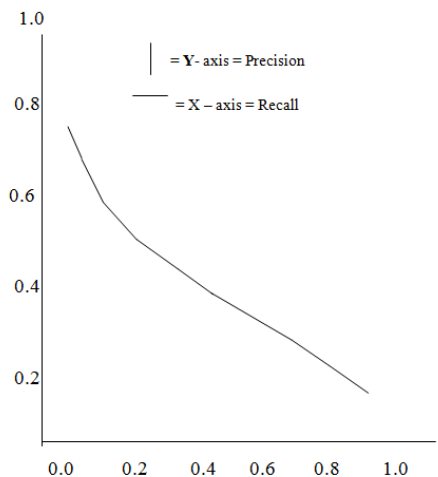

Figure 4. Optimal Precision /Recall Graph

## 2. INFORMATION RETRIEVAL AND WEB SEARCH

Information Retrieval (IR) is the area of computer science concerned with retrieving information about a subject from a collection of data objects. This is not the same as Data Retrieval, which in the context of documents consists mainly in determining which documents of collections contain the keywords of a user query. Information Retrieval deals with satisfying a user need.

 Although there was an important body of Information Retrieval techniques published before the invention of the World Wide Web, there are unique characteristics of the Web that made them unsuitable or insufficient.

This idea is also present in a survey about Web search by Brooks [3], which states that a distinction could be made between the closed web, which comprises high-quality controlled collections on which a search engine can fully trust and the open web, which includes the vast majority of Web pages and on which traditional IR techniques concepts and methods are challenged.

The advent of the World Wide Web caused a dramatic increase in the usage of the Internet. The World Wide Web is a broadcast medium where a wide range of information can be obtained at a low cost. Information on the World Wide Web is important not only to individual users, but also to the organizations especially when the critical decision making is concerned. Most users obtain World Wide Web information using a combination of search engines [4] and browsers. However these two types of retrieval mechanisms do not necessarily produce the all information needed by the user. Hundreds of irrelevant documents returned in response to a search query, only less than 18% of web pages are relevant to the user. To overcome this problem we need one effective search engine, which produces maximum relevant information with minimum time and low cost.

## 3. WEB CRAWLER

A Web crawler is a computer program that browses the World Wide Web in a methodical, automated manner or in an orderly fashion. Web crawlers are mainly used to create a copy of all the visited pages for later processing by a search engine that will index the downloaded pages to provide fast searches. Crawlers can also be used for automating maintenance tasks on a Web site, such as checking links or validating html code [5]. Crawlers can be used to gather specific types of information from Web pages, such as harvesting e-mail addresses (usually for sending spam).web search engines are becoming increasingly, important as the primary means of locating relevant information.

A good crawler for a large search engine has to deal with two issues [7 , 8]. First, it has to have a good crawling strategy, i.e., an excellent strategy for deciding which pages to download next. Secondly, it desires to have an extremely optimized system architecture that can download a large number of pages per second while being healthy and strong against crashes, manageable, and considerate of resources and web servers.

We need to implement the system that considers the performance of a focused crawler built on top of a general-purpose database system, although the throughput of that system is still significantly below that of a high-performance bulk crawler. However, in the case of a larger search engine, we need to combine good crawling strategy and optimized system design. The number of possible crawlable URLs being generated by server-side software has also made it difficult for web crawlers to avoid retrieving duplicate content. Endless combinations of HTTP get (URL-based) parameters exist, of which only a small selection will actually return unique content. For example, a simple online photo gallery may offer three options to users, as specified through HTTP GET parameters in the URL. If there exist four ways to sort images, three choices of thumbnail size, two file formats, and an option to disable user-provided content, then the same set of content can be accessed with 48 different URLs, all of which may be linked on the site. This mathematical combination creates a problem for crawlers, as they must sort through endless combinations of relatively minor scripted changes in order to retrieve unique content.

# 4. BASIC WEB CRAWLER ARCHITECTURE

A Web Crawler must be developed by consideration of different applications and polices with a reasonable amount of work. Note that there are significant differences between the scenarios. For example, a breadth-first crawler [9] has to keep track of which pages have been crawled already; this is commonly done using a "URL seen" data structure that may have to reside on disk for large crawls.

Web crawler is also called as software agent. In general, it starts with a list of URLs [10] to visit, called the seeds. As the crawler visits these URLs, it identifies all the hyperlinks in the page and adds them to the list of URLs to visit, called the crawl frontier. A crawler must not only have a good crawling strategy, as mentioned earlier, but it should also have a highly optimized architecture. The large volume implies that the crawler can only download a fraction of the Web pages within a given time, so it needs to prioritize it's downloads. The high rate of change implies that the pages might have already been updated or even deleted. [11]Web crawlers typically identify themselves to a Web server by using the User-agent field of an HTTP request. It is important for web crawlers to identify themselves so that web site administrators can contact the owner if needed. It is important for Web crawlers to identify themselves so that Web site administrators can contact the owner if needed. In some cases, crawlers may be accidentally trapped in a crawler trap or they may be overloading a Web server with requests, and the owner needs to stop the crawler. Identification is also useful for administrators that are interested in knowing when they may expect their Web pages to be indexed by particular search engine affiliations.

# 5. EXAMPLES OF WEB CRAWLER

## 5.1. Google Search

Google Web Search is a web search engine owned by Google Inc [12]. Google Search is the most-used search engine on the World Wide Web receiving several hundred million queries each day through its various services [13]. The order of search results on Google's search-results pages is based, in part, on a priority rank called a "Page Rank" [14]. Google Search provides many options for customized search, using Boolean operators such as: exclusion ("-xx"), inclusion ("+xx"), alternatives ("xx OR yy"), and wildcard ("x * x"). The main purpose of Google Search is to hunt for text in Web pages, as opposed to other data, such as with Google. Google Search was originally developed by Larry Page and Sergey Brin in 1997. In figure 5. we have shown the estimated size of Google's index for the past 6 months.

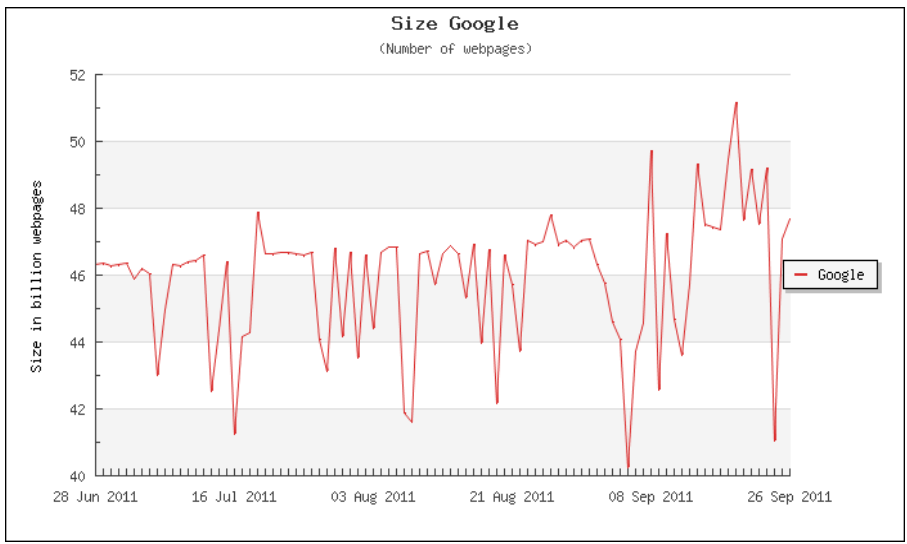

Figure 5. The Estimated size of Google's index

## 5.2. Bing Search

Bing (formerly Live Search, Windows Live Search, and MSN Search) is a web search engine [15] (advertised as a "decision engine" [16]) from Microsoft. Bing was unveiled by Microsoft CEO Steve Ballmer on May 28, 2009 at the All Things Digital conference in San Diego. It went fully online on June 3, 2009, with a preview version released on June 1, 2009.

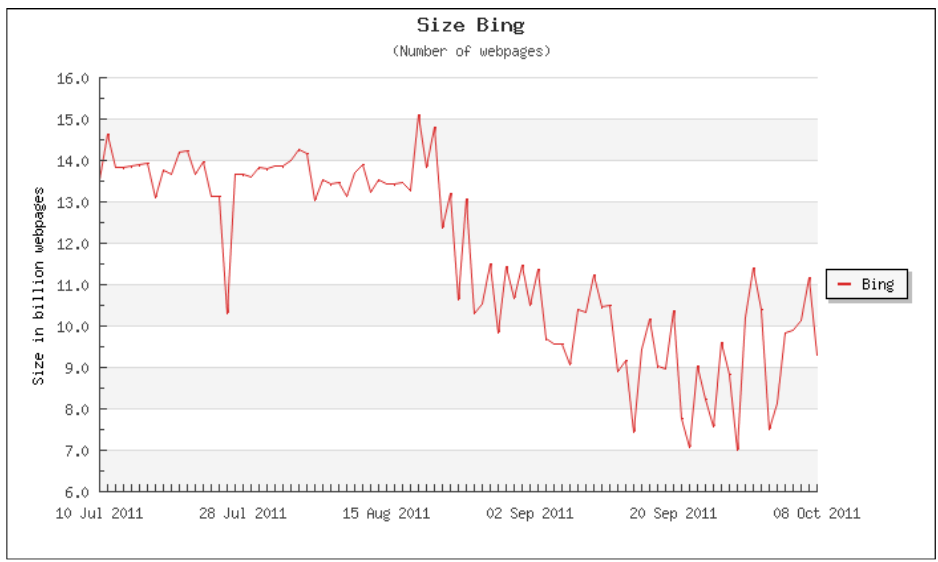

Figure 6. The Estimated size of Bing index

Notable changes include the listing of search suggestions as queries are entered and a list of related searches (called "Explore pane") based on semantic technology [17] from Power set that Microsoft purchased in 2008. On July 29, 2009, Microsoft and Yahoo! announced a deal in which Bing would power Yahoo. All Yahoo! Search global customers and partners are expected to have made the transition by early 2012. In figure 6. we have shown the estimated size of Bing index for the past 6 months.

## 6. PROPOSED EPOW ARCHITECTURE

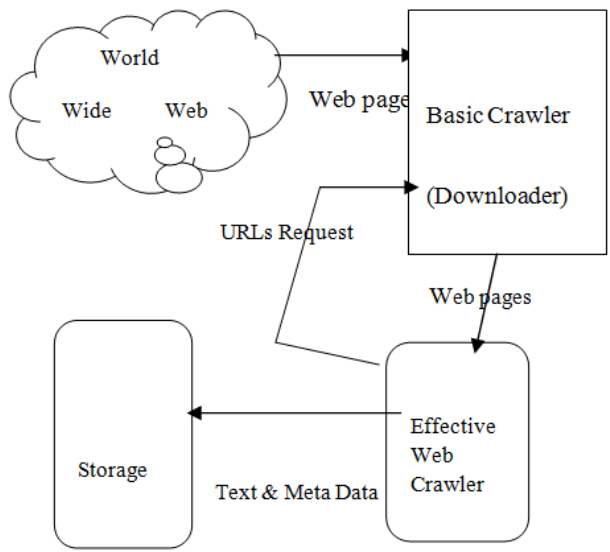

Figure.7  Effective Performance of Web Crawler [EPOW] Architecture

In our EPOW (Effective Performance of WebCrawler) approach, we have designed one Basic Crawler and Effective web crawler (i.e. master crawler). Basic Crawler consists of multiple downloaders. It fetches web pages (documents) from the World Wide Web when provided with a corresponding URL. Effective web crawler receives the URLs which is sent by the basic crawler and stored in the priority Queue. On priority bases it analyses the request and issues a new request to the basic crawler (Downloader).  The basic crawler downloads the requested pages and supplies them to the master crawler for storage and analysis. This process will be iteratively done till maximum relevant documents are fetched from the web at regular interval.

In EPOW (Effective Performance of Web Crawler) method to improve the performance we have personalized the parallelization policy. The aim of using this policy is to maximize the download rate while minimizing the overhead from parallelization.   In this method we have included multiple downloaders in the basic crawler, and to receive multiple requests we have used circular queue and priority queue in the effective web crawler. All URLs are added in to circular queue, from which they are extracted in a certain order and corresponding pages are downloaded from the web.URL extraction and en-queuing steps are repeated on the newly found content in an iterative manner. After de-queuing the each request, Master crawler analyzes the document and sends multiple URLs list which is relevant to the previous document. We need to stop the retrieval of web pages at certain interval, Due to the large size of Web. For that purpose we have proposed one scheduler in our Effective Web Crawler.

The web has a very dynamic nature and crawling a fraction of the web can take weeks or months. By the time a web crawler has finished its crawl, many events can have happened, including creations, updates and deletions. The pages will remain outdated due to these modifications. We have also considered this problem in our system. The objective of our EPOW crawler is to keep the average freshness [18, 19] of pages in its collection as high as possible, or to keep the average age of pages as low as possible. For that we have adapted optimal Revisit Policy .This method is for keeping average freshness high by ignoring the pages that change too often, and the optimal for keeping average age low is to use access frequencies that monotonically increase with the rate of change of each page.

# 7. ADVANTAGES

The following are the advantages of an Effective Web crawler.

## 7.1 Flexibility &Portability

 As earlier discussed, we would like to be able to use the system in a variety of scenarios, across the various platforms with as few modifications if required. Our system provides flexibility & portability.

## 7.2 Low Cost and High Performance

 The system should scale to at least several hundred pages per second and hundreds of millions of pages per run, and should run on low-cost hardware. Note that efficient use of disk access is crucial to maintain a high speed after the main data structures. Our system will provide high performance with low cost.

## 7.3 Robustness

 There are several aspects here. First, since the system will interact with millions of servers, it has to tolerate bad HTML, strange server behaviour and configurations, and many other odd issues. Our goal is, if the system gets it wrong, and then it prompts the message with caution, and if necessary ignores pages and even entire servers with odd behaviour, since in many applications we can only download a subset of the pages anyway. Secondly, since a crawl may take weeks or months, the system needs to be able to tolerate crashes and network interruptions without losing (too much of) the data. Thus, the state of the system needs to be kept on disk. We note that we do not really require strict ACID properties. Instead, we decided to periodically synchronize the main structures to disk, and to recrawl a limited number of pages after a crash.

## 7.4 Speed Control

We have to avoid putting too much load on a single server; we do this by contacting each site only once every 20 second unless specified otherwise. It is also desirable to throttle the speed on a domain level, in order not to overload small domains. Finally, since we are in a campus environment where our connection is shared with many other users, we also need to control the total download rate of our crawler. In particular, we crawl at low speed during the peak usage hours of the day, and at a much higher speed during the late night and early morning, limited mainly by the load tolerated by our main campus router.

# 8. CONCLUSION AND FUTURE DIRECTIONS

In this paper we have described the necessity of Information Retrieval System and defined two major measures commonly associated with information systems are precision and recall. We have proposed a new system to increase the precision of information retrieval and organization retrieval results. In our paper we have also compared Information Retrieval with web searching process. We have addressed the architecture details of our EPOW (Effective Performance of Web

Crawling) system for searching the relevant documents with less amount of time. In our system to maximize the download rate while minimizing the overhead we have considered multiple downloaders in the basic crawler. We have also discussed how to handle multiple requests from the basic crawler by using circular queue. We have proposed scheduler mechanism in our effective web crawler to stop the fetching the relevant web pages at certain time interval.

Mainly, Web has very dynamic nature; due to this most of the pages will be outdated at certain period. To reduce this problem we have used optimal revisit policy. The objective of this method is to keep the average freshness of pages as high as possible and for keeping average age of pages as low as possible. Freshness that indicates whether the local copy is accurate or not. Whereas age that indicates how outdated the local copy is. The Revisiting policies considered in our system regard all pages as homogeneous in terms of quality i.e. all the pages on the web are worth the same. But in realistic it is quite different. So we are looking further information about the web page quality should be included to achieve an Excellency.

## ACKNOWLEDGEMENTS

We would like to thank every one, who has motivated and supported us for preparing this Manuscript.

**Authors**

Mr.Shaik.Abdul Nabi is the Head of the Dept. of CSE, AVN Inst.Of Engg.Tech, Hyderabad, AP, and India. He completed his B.E (Computer Science) from Osmania University, A.P. He received his M.Tech. From JNTU Hyderabad campus and currently he is pursuing Ph.D. in the area of Web Mining from Acharya Nagarjuna University, Guntur, AP, and India. He is a certified professional by Microsoft. His expertise areas are Data warehousing and Data Mining, Data Structures & UNIX Networking Programming.

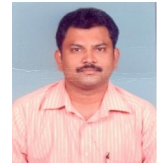

Dr P.Premchand is a professor in department of Computer Science & Engineering, Osmania University, Hyderabad, A.P, and India. He completed his ME (Computer Science) from Andhra University, A.P. He has received Ph.D degree from Andhra University, A.P. He guided many scholars towards the award of Ph.D degree from various Universities. He was a director of AICTE, New Delhi, during 1998-99. He also worked as Head of the Dept of CSE and Additional Controller of Examinations, Osmania University, AP. Now currently he is a chairman of BOS and Dean for Faculty of Engineering, Osmania University, AP., India.

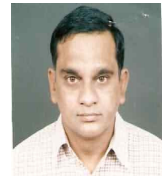